\documentclass{eptcs}
\providecommand{\event}{ACL2 2015} 
\usepackage{url}
\usepackage{breakurl}
\usepackage{graphicx}
\title{Reasoning About LLVM Code Using Codewalker}
\author{
David S. Hardin\\
\institute{Advanced Technology Center\\Rockwell Collins\\
Cedar Rapids, IA, USA}
\email{david.hardin@rockwellcollins.com}
}

\def\titlerunning{Reasoning About LLVM Code Using Codewalker}
\def\authorrunning{D. Hardin}

\begin{document}
\maketitle

\begin{abstract}

This paper reports on initial experiments using J Moore's Codewalker 
to reason about programs compiled to the Low-Level Virtual Machine (LLVM)
intermediate form.  Previously, we reported on a translator from LLVM to the
applicative subset of Common Lisp accepted by the ACL2 theorem prover,
producing executable ACL2 formal models, and allowing us to both prove
theorems about the translated models as well as validate those models by testing.
That translator provided many of the benefits of a pure decompilation
into logic approach, but had the disadvantage of not being verified.
The availability of Codewalker as of ACL2 7.0 has provided an opportunity to 
revisit this idea, and employ a more trustworthy decompilation into logic tool.
Thus, we have employed the Codewalker method to create an interpreter 
for a subset of the LLVM instruction set, and have used Codewalker to 
analyze some simple array-based C programs compiled to LLVM form.
We discuss advantages and limitations of the Codewalker-based method
compared to the previous method, and provide some challenge problems
for future Codewalker development.

\end{abstract}


\section{Introduction}\label{intro}

In previous work~\cite{HardinACL214}~\cite{LLVMtoACL2}, we 
built a translator from Low-Level Virtual Machine (LLVM) intermediate
form \cite{LLVM} to the applicative subset of Common Lisp
\cite{CommonLispHyperSpec} accepted by the ACL2 theorem prover
\cite{ACL2book}, and performed verification on the translated form using
ACL2's automated reasoning capabilities.

LLVM is the intermediate form for many common compilers, including the
\texttt{clang} compiler used by Apple OS X and iOS developers.  LLVM
supports a number of language frontends, and LLVM code generation
targets exist for a wide variety of machines, including both CPUs and
GPUs.  LLVM is a register-based intermediate language in Static Single
Assignment (SSA) form \cite{SSA}.  As such, LLVM supports any number
of registers, each of which is only assigned once, statically
(dynamically, of course, a given register can be assigned any number
of times).  Andrew Appel has observed that ``SSA form is a kind of functional
programming'' \cite{SSAfun}; this observation, in turn, inspired us to
build a translator from LLVM to the applicative subset of Common Lisp
accepted by the ACL2 theorem prover.  Our translator,
written in OCaml \cite{OCaml}, produced an
executable ACL2 specification that was able to support proof-based 
verification, as well as validation via testing.

The above approach was satisfactory for the technology that we had at 
hand for use with ACL2 in 2013, but had the obvious weakness of relying on a fair amount
of unverified code.  The situation changed in late 2014, when J Moore 
released the initial version of Codewalker, an instruction-set-neutral
decompilation-into-logic system, with ACL2 7.0~\cite{Codewalker}.   
Thus, an experiment began in early 2015 
to determine whether Codewalker could be used to produce a similar proof
environment for LLVM code.

\section{An Example}\label{example}

As an example, consider the C source code of Figure \ref{Ccode}.
This function counts the number of occurrences of a given value in the
first n elements of an array.  (NB: By default the \texttt{clang} compiler
treats all \texttt{int} values as 32 bits wide, and all \texttt{long}
values as 64 bits wide.)

\begin{figure*}
\begin{verbatim}
unsigned long occurrences(unsigned long val, unsigned int n, 
                          unsigned long *array) {
  unsigned long num_occur = 0;
  unsigned int j = 0;
  for (j = 0; j < n; j++) {
    if (array[j] == val) num_occur++;
  }
  return num_occur;
}
\end{verbatim}
\hrulefill
\caption{Example C code to count occurrences of an input value in an array.}
\label{Ccode}
\end{figure*}

This is an admittedly simple example, but it 
allows us to narrate a complete analysis within 
the confines of this paper, and should be within Codewalker's 
capabilities to analyze.  We have also performed similar analyses for
other small C programs, namely tail-recursive factorial, as well as  
a program to compute the sum of array elements.

LLVM code for this function is produced by invoking \texttt{clang}
as follows: \texttt{clang -O1 -S -emit-llvm occurrences.c}.  The
generated LLVM code for clang version 6.1.0 (which supports LLVM 3.6.0) is
excerpted in Figure \ref{LLVMcode}; this is essentially the same code
as reported in~\cite{HardinACL214}.

\begin{figure*}
\begin{verbatim}
define i64 @occurrences(i64 %val, i32 %n, i64* %array) {
  %1 = icmp eq i32 %n, 0
  br i1 %1, label %._crit_edge, label %.lr.ph

.lr.ph:
  %indvars.iv = phi i64 [ %indvars.iv.next, %.lr.ph ], [ 0, %0 ]
  %num_occur.01 = phi i64 [ %.num_occur.0, %.lr.ph ], [ 0, %0 ]
  %2 = getelementptr inbounds i64* %array, i64 %indvars.iv
  %3 = load i64* %2, align 8, !tbaa !1
  %4 = icmp eq i64 %3, %val
  %5 = zext i1 %4 to i64
  %.num_occur.0 = add i64 %5, %num_occur.01
  %indvars.iv.next = add nuw nsw i64 %indvars.iv, 1
  %lftr.wideiv = trunc i64 %indvars.iv.next to i32
  %exitcond = icmp eq i32 %lftr.wideiv, %n
  br i1 %exitcond, label %._crit_edge, label %.lr.ph

._crit_edge:
  %num_occur.0.lcssa = phi i64 [ 0, %0 ], [ %.num_occur.0, %.lr.ph ]
  ret i64 %num_occur.0.lcssa
}
\end{verbatim}
\hrulefill
\caption{LLVM code for the occurrences example.}
\label{LLVMcode}
\end{figure*}

Observe that LLVM output is similar to assembly code, with labels and
low-level opcodes like \texttt{br} (branch), \texttt{icmp} (integer
compare) and \texttt{load} (load from memory).  Registers are
prepended with the ``\%'' character, and are given
sometimes-meaningful names.  Consistent with the SSA philosophy, no
register appears on the left hand side of an assignment (``='') more
than once.  A peculiar feature of LLVM code is the \texttt{phi}
instruction, which provides register renaming at a branch target.

\subsection{Translation to ACL2 Syntax}\label{translation}

In previous work, we automatically translated the above LLVM program 
into an ACL2 functional program.  In the current work, we merely translate
the LLVM assembly code syntax into a form that is easier for ACL2 to process.
The translated form for the LLVM code 
of Figure \ref{LLVMcode} is depicted in Figure \ref{LLVMcodeACL2}. 

\begin{figure*}
\begin{verbatim}
    ;;    reg[2] contains val
    ;;    reg[1] contains n
    ;;    reg[0] contains array base address


    (CONST 0)           ; 0
    (POPTO 3)           ; 1   reg[3] <- 0
    (EQ 4 1 3)          ; 2   n == 0?

    (CONST 0)           ; 3
    (POPTO 5)           ; 4   phi(j), j <- 0
    (CONST 0)           ; 5
    (POPTO 6)           ; 6   phi(num_occur), num_occur <- 0

    (BR 4 14 1)         ; 7   branch to ._crit_edge if n == 0

;; .lr.ph:
    (GETELPTR 7 0 5)    ; 8   reg[7] <- mem address of arr[index]
    (LOAD 8 7)          ; 9   reg[8] <- mem[reg[7]] = arr[index]
    (EQ 9 8 2)          ; 10  reg[8] == val?
    (ADD 10 6 9)        ; 11  num_occur conditional increment
    (CONST 1)           ; 12
    (POPTO 11)          ; 13
    (ADD 12 5 11)       ; 14  reg[12] <- j+1
    (EQ 13 12 1)        ; 15  j+1 == n?

    (PUSH 12)           ; 16
    (POPTO 5)           ; 17  phi(j), j <- j+1
    (PUSH 10)           ; 18
    (POPTO 6)           ; 19  phi(num_occur)

    (BR 13 1 -12)       ; 20  loop back to .lr.ph if j+1 < n

;;._crit_edge:
    (PUSH 6)            ; 21  push num_occur on stack
    (HALT)              ; 22
\end{verbatim}
\hrulefill
\caption{ACL2 representation of the LLVM code for the occurrences example.}
\label{LLVMcodeACL2}
\end{figure*}

The instruction format is straightforward: if the LLVM instruction is 
\texttt{a = ins b c}, then the ACL2 syntax is \texttt{(INS A B C)}.  
Thus, \texttt{(ADD x y z)} stores the sum of the contents of registers 
(locals) y and z in register x; and  \texttt{(BR E F G)} branches to 
the instruction word at the current program counter + offset F 
if register E is nonzero, and to the instruction word at the 
current program counter + offset G otherwise.  A few new 
instructions have been added to aid in phi processing: \texttt{(CONST X)} 
pushes a constant value X on a LIFO stack; \texttt{(PUSH Y)} pushes 
the contents of register Y onto the stack; and \texttt{(POPTO Z)} pops 
the top of stack value into register Z.    We also define 
a \texttt{(HALT)} instruction so we don't have to worry 
about defining a return linkage (this is future work).

Each instruction occupies one instruction word (of indeterminate size), 
and each register holds an unbounded integer.  This represents a 
slight loss of fidelity relative to the previous work, but we 
thought it unwise to tackle issues related to both
Codewalker and modular arithmetic at the same time. 

\section{LL2: An LLVM Subset Interpreter}\label{interp}

Before being able to utilize Codewalker, we must first 
define an operational semantics, or interpreter, for the target instruction set.  
The Codewalker sources provide one such
example interpreter, for the M1 subset of the Java Virtual 
Machine (JVM)~\cite{JVMSpec}.  We used this ACL2 code as 
the basis for our LLVM subset interpreter, called LL2.  As is 
typical with such an interpreter written in ACL2, a machine state data
structure is declared, and passed as a parameter to all functions that
read and/or write elements of the state.  If a given function updates
the state, the modified state must be returned.  Obviously, for a large
state, functional update of the state can become quite expensive.  
Thus, an ACL2 single-threaded object (stobj) \cite{STOBJ} is often 
used to represent state.  The destructive update property of stobjs 
provides good performance when executing functions on 
concrete state.  The LL2 machine stobj, called simply \texttt{s}, 
contains fields for the Program Counter (PC), 
local variables, memory, stack, and program storage.  All but the
first can be thought of as lists.  Accessor and updater functions 
are defined for all fields, with updaters preceded by a `!' character; 
thus \texttt{(loi k s)} retrieves the kth local variable (or register,
in LLVM parlance), while \texttt{(!loi j val s))} updates the value of 
the jth register to \texttt{val}.  Note that \texttt{(loi k s)}
is defined as \texttt{(nth k (rd :locals s))}, and \texttt{(!loi j val
  s)} is defined as \texttt{(wr :locals (update-nth j val (rd :locals s)) s)}.

Once the machine state data structure is defined, semantic functions 
need to be written for all supported instructions.  For example, the
semantic function for \texttt{(EQ x y z)} is as follows:

\begin{verbatim}
(defun execute-EQ (inst s)
  (declare (xargs :stobjs (s)))
  (let* ((s (!loi (arg1 inst)
                  (if (= (loi (arg2 inst) s) (loi (arg3 inst) s)) 1 0) s))
         (s (!pc (+ 1 (pc s)) s)))
    s))
\end{verbatim}

where \texttt{inst} is the list form of an instruction (as depicted in
Figure~\ref{LLVMcodeACL2}), \texttt{(arg1 inst)}
is \texttt{(nth 1 inst)},  
\texttt{(arg2 inst)} is \texttt{(nth 2 inst)}, and 
\texttt{(arg3 inst)} is \texttt{(nth 3 inst)}.  Thus,
\texttt{execute-EQ} stores the value 1 in the register indicated by the 
first argument if the value stored in the register indicated by 
the second argument is equal to the value stored in the register indicated by 
the third argument; the value 0 is stored in the first argument register
otherwise.  Finally, the program counter is incremented. 

%

Once semantic functions have been written for every supported instruction,
a simple instruction selector function can
be composed, as follows:

\begin{verbatim}
(defun do-inst (inst s)
  (declare (xargs :stobjs (s)))
  (if (equal (op-code inst) 'ADD)
      (execute-ADD inst s)
    (if (equal (op-code inst) 'BR)
        (execute-BR inst s)
      (if (equal (op-code inst) 'CONST)
          (execute-CONST inst s)
          ... s)))...)
\end{verbatim}

This instruction selector function is called by the instruction
stepper function:

\begin{verbatim}
(defun step (s)
  (declare (xargs :stobjs (s)))
    (let ((s (do-inst (next-inst s) s)))
      s))
\end{verbatim}

where \texttt{(next-inst s)} is \texttt{(nth (pc s) (program s))}.

Finally, the instruction stepper is called by the top-level LL2 interpeter:

\begin{verbatim}
(defun ll2 (s n)
  (declare (xargs :stobjs (s)))
  (if (zp n)
      s
      (let* ((s (step s)))
        (ll2 s (- n 1)))))
\end{verbatim}

Note that this is all fairly standard technique for defining an 
instruction set interpreter in ACL2; one peculiarity, however, is that the
top-level interpreter argument order (namely, state followed by step count) 
is mandated by Codewalker.

\subsection{Concrete Execution}\label{exec}

It is advantageous to be able to validate LLVM programs by
running them against concrete inputs.  Since all of our interpreter functions are
executable, we can readily perform such validation testing.  In the ACL2 code
of Figure \ref{concrete}, we set up an initial state, establishing an
array of length 8 starting at address 100.  We write
various values into memory at increasing addresses.  The array base
address is stored in local 0, followed by the
\texttt{n} and \texttt{val} parameters, in locals 1 and 2,
respectively.  The program is written using the \texttt{(wr :program
  '(...))} form.  The program is stepped to conclusion by invoking 
\texttt{(ll2 s 113)}; the return value can be found at \texttt{(loi 6 s)}.

\begin{figure*}
\begin{verbatim}
(include-book "LL2")
(in-package "LL2")

(!loi 0 100 s)
(!loi 1 8 s)
(!loi 2 399 s)
(!memi 100 399 s)
(!memi 101 234 s)
(!memi 102 0 s)
(!memi 103 75 s)
(!memi 104 399 s)
(!memi 105 399 s)
(!memi 106 (1- (expt 2 64)) s)
(!memi 107 20 s)
(!pc 0 s)

(wr :program '((CONST 0)...))

(ll2 s 113)   ;; run to HALT
\end{verbatim}
\hrulefill
\caption{Concrete test case for the occurrences example.}
\label{concrete}
\end{figure*}

As we have written the value 399 into the array three times, when we
run the interpreter and fetch the return result as described above, we 
obtain the expected value: 3.  The interpreter executes approximately 
226,000 LLVM instructions per second on an ordinary laptop computer.  
This is approximately one-tenth the speed of our previous method, 
as is to be expected for an interpreted vs. compiled approach, but 
this performance level is still more than adequate for validation testing.

\section{Codewalker}

Now that the interpreter for LL2 is in place, we can begin to use
Codewalker to perform decompilation into logic for LLVM programs, 
producing semantic functions for those programs that 
the ACL2 user can further reason about.  The end goal is to prove that
the LLVM code for a given function implements a much more abstract
function, written in ACL2, about which we can readily prove interesting
correctness properties.  In the extensive code
documentation for Codewalker, the system is described as follows~\cite{Codewalker}:

\begin{quote}
Two main facilities are provided by Codewalker: the abstraction of a piece of
code into an ACL2 ``semantic function'' that returns the same machine state,
and the ``projection'' of such a function into another function that computes
the final value of a given state component using only the values of the
relevant initial state components.

Codewalker is independent of any particular machine model, as long as a
step-based operational semantics for the machine is defined in ACL2.  To
facilitate this language-independent analysis, the user must declare a
``model API'' that allows Codewalker to access functionality of the model
(e.g., setting the pc in a symbolic state).  Generally speaking, Codewalker
accesses the model by forming symbolic ACL2 expressions that answer certain
questions, then applying the ACL2 simplifier with full access to user-proved
lemmas, and then inspecting the resulting term to recover the answer.
\end{quote}

Thus, to begin, we tell Codewalker about our operational semantics 
using \texttt{def-model-api}, telling it the name of our interpreter
function, the state variable, whether the state is a stobj, the name
of the step function, and so on.  We next introduce the program to 
be analyzed, and prove some simple theorems 
about it, e.g. that writes to state fields other than the 
program field don't affect the program.

Next, we provide Codewalker with important program-level invariants as 
well as loop invariants.  We also assist the system by providing a
measure for the loop clock function, as illustrated in Figure \ref{invariants}.

\begin{figure*}
\begin{verbatim}
(defun hyps (s)
   (declare (xargs :stobjs (s)))
   (and (sp s)
        (natp (rd :pc s))
        (< (rd :pc s) (len (rd :program s)))
        (< 16 (len (rd :locals s)))
        (integer-listp (rd :locals s))
        (integer-listp (rd :memory s))
        (integer-listp (rd :stack s))))

(defun-nx loop-pc-p (s)
  (= 8 (rd :pc s)))

(defun-nx loop-inv (s)
  (< (nth 5 (rd :locals s))
     (nth 1 (rd :locals s))))

(defun-nx program-inv (s)
  (and (natp (nth 0 (rd :locals s)))
       (natp (nth 1 (rd :locals s)))
       (integerp (nth 2 (rd :locals s)))
       (natp (nth 3 (rd :locals s)))
       (natp (nth 5 (rd :locals s)))
       (natp (nth 6 (rd :locals s)))))

(defun-nx clk-8-measure (s)
  (nfix (if (not (loop-pc-p s))
            (nth 1 (rd :locals s))
          (- (nth 1 (rd :locals s))
             (nth 5 (rd :locals s))))))
\end{verbatim}
\hrulefill
\caption{Some invariants and measures provided to Codewalker.}
\label{invariants}
\end{figure*}

Finally, we set Codewalker to work, by invoking its
\texttt{def-semantics}  function.  First, we ask Codewalker to 
generate a semantic function for the ``preamble'' of the code 
(before the loop), then ask it to produce a semantic function 
for the loop itself, as shown in Figure \ref{def-semantics}.  
We often wish to break up the processing in this way, and not 
give the entire function to Codewalker in a single chunk.  One reason 
for this is that it can be tricky to craft just the right invariants that are 
true for preamble, as well as the loop and postlude, and that 
Codewalker will be able to process successfully.  
  
Codewalker development is still in its early phase, and the system is a bit 
``touchy'' when it comes to the combination of 
focus regions, invariants, measure annotations, and so on 
that will result in success.  In Codewalker's defense, it is 
very sophisticated software attempting a very difficult job.  
To quote the Codewalker documentation: 
``Def-semantics actually prints a lot of stuff as it goes.  It also often
fails!  Some of its error messages make supposedly helpful suggestions as to
what's `wrong.'  Often your response will be to prove more lemmas because
things aren't being reduced to the canonical forms.  Another response
might be to restrict the focus region or strengthen the invariant so as to
avoid certain cases.''~\cite{Codewalker}

\begin{figure*}
\begin{verbatim}
(def-semantics
  :init-pc 0
  :focus-regionp (lambda (pc) (and (<= 0 pc) (< pc 8)))
  :root-name preamble
  :hyps+ ((occurrences-programp s)
          (program-inv s)))

(def-semantics
  :init-pc 8
  :focus-regionp (lambda (pc) (>= pc 8))
  :root-name loop
  :hyps+ ((occurrences-programp s)
          (loop-inv s) (program-inv s)
          (<= (+ (nth 0 (rd :locals s)) (nth 1 (rd :locals s))) 
              (len (rd :memory s))))
  :annotations ((clk-loop-8 (declare (xargs :measure (clk-8-measure s))))
                (sem-loop-8 (declare (xargs :measure (clk-8-measure s))))))
\end{verbatim}
\hrulefill
\caption{Invocations of Codewalker \texttt{def-semantics} for the
  occurrences example.}
\label{def-semantics}
\end{figure*}

Codewalker produces decompilations of the indicated code segments, 
which we can then assemble using functional composition, e.g.:

\begin{verbatim}
(defun-nx composition (s)
  (sem-loop-8 (sem-preamble-0 s)))
\end{verbatim}

Codewalker also produces correctness theorems about the 
generated semantics functions, e.g.:

\begin{verbatim}
(DEFTHM SEM-PREAMBLE-0-CORRECT
  (IMPLIES (AND (HYPS S)
                (OCCURRENCES-PROGRAMP S)
                (PROGRAM-INV S)
                (EQUAL (RD :PC S) 0))
           (EQUAL (LL2 S (CLK-PREAMBLE-0 S))
                  (SEM-PREAMBLE-0 S))))

(DEFTHM SEM-LOOP-8-CORRECT
  (IMPLIES (AND (HYPS S)
                (OCCURRENCES-PROGRAMP S)
                (LOOP-INV S)
                (PROGRAM-INV S)
                (<= (+ (NTH 0 (RD :LOCALS S))
                       (NTH 1 (RD :LOCALS S)))
                    (LEN (RD :MEMORY S)))
                (EQUAL (RD :PC S) 8))
  (EQUAL (LL2 S (CLK-LOOP-8 S))
         (SEM-LOOP-8 S))))
\end{verbatim}

The latter theorem states that if the LL2 interpreter is poised at 
the top of the loop (pc = 8) then running the LL2 interpreter with the
occurrences program loaded for a proper
number of steps (given by \texttt{(CLK-LOOP-8 S)} yields the 
same result as executing the generated semantic function.

\section{Reasoning about LLVM Code via Codewalker Semantic Functions}\label{reasoning}

In order to reason about a function such as \texttt{occurrences} in
ACL2, we first need to perform abstraction on the data types;
particularly, we wish to abstract the input array to a Lisp list.
Since we are utilizing stobjs, however, this abstraction has already
been provided for us.  (Recall that stobjs provide a list abstraction
for array data types that feature an efficient, in-place, destructive implementation.)

Next, we need a ``golden'' list-based specification of
\texttt{occurrences}.  This function should be easy to reason
about using ACL2, and so should be written in non-tail-recursive style,
as in the following:

\begin{verbatim}
(defun occurlist (val lst)
  (declare (xargs :guard (and (integerp val) (integer-listp lst))))
  (if (endp lst)
      0
    (+ (if (= val (car lst)) 1 0)
       (occurlist val (cdr lst)))))
\end{verbatim}

We wish to prove that the execution of the LLVM instructions 
of the compiled \texttt{occurrences} function
operating over an array in memory produces a result equal to the
\texttt{occurlist} function operating over a list.  Unfortunately for
the proof of the above, the semantic functions
generated by Codewalker are tail-recursive.  The proof actually
proceeds by the use of two additional functions, a pair of 
tail-recursive/non-tail-recursive functions that are generated and 
proved equal by \texttt{defiteration}, a book found in 
\texttt{centaur/misc} in the standard ACL2 distribution.  
(This technique was earlier described in~\cite{ACL2GPU}.)  
The call to \texttt{defiteration} is as follows:

\begin{verbatim}
(acl2::defiteration occur-arr (num val s)
  (declare (xargs :stobjs s
                  :guard (and (integerp num) (integerp val))))
  (ifix (+ (if (= (nth ix (rd :memory s)) val) 1 0) num))
   :returns num
   :index ix
   :last (len (rd :memory s)))
\end{verbatim}

We first prove that the value stored in the \texttt{num\_occur}
register (register 6) after execution of the composition of 
semantic functions generated by Codewalker is equal to the 
result of the tail-recursive function generated by the call 
to \texttt{defiteration} above:

\begin{verbatim}
(defthm composition-=-occur-arr-tailrec
  (implies
   (and (hyps s)
        (program-inv s)
        (occurrences-programp s)
        (<= (+ (nth 0 (rd :locals s)) (nth 1 (rd :locals s))) 
            (len (rd :memory s)))
        (= (nth 1 (rd :locals s)) (len (rd :memory s))))
   (= (nth 6 (rd :locals (sem-loop-8 (sem-preamble-0 s))))
      (occur-arr-tailrec 0 0 (nth 2 (rd :locals s)) s)))
  :hints (("Goal" :in-theory (enable occur-arr-tailrec)
           :cases ((= (len (rd :memory s)) 0) (> (len (rd :memory s)) 0)))))
\end{verbatim}

We then prove that the non-tail-recursive function generated by
\texttt{defiteration} is equal to occurlist:

\begin{verbatim}
(defthm occur-arr-iter-=-occurlist
  (implies
   (and (sp s) (integerp val) (integer-listp (rd :memory s))
        (= (len (rd :memory s)) (len (rd :memory s))))
   (= (occur-arr-iter (len (rd :memory s)) 0 val s)
      (occurlist val (rd :memory s)))))
\end{verbatim}

The above theorem can be proved by first proving the following lemma:

\begin{verbatim}
(defthm occur-arr-iter-=-occurlist-take--thm
  (implies
   (and
    (sp s) (natp xx) (integerp val) 
    (integer-listp (rd :memory s))
    (<= xx (len (rd :memory s))))
   (= (occur-arr-iter xx 0 val s)
      (occurlist val (take xx (rd :memory s)))))
  :hints (("Subgoal *1/1" :in-theory (enable occur-arr-iter))))
\end{verbatim}

Since \texttt{occur-arr-iter} and \texttt{occur-arr-tailrec} are
already proved equal by \texttt{defiteration}, the proof of 
\texttt{composition-=-occurlist} then follows readily.

\begin{verbatim}
(defthm composition-=-occurlist
  (implies
   (and (hyps s)
        (program-inv s)
        (occurrences-programp s)
        (<= (+ (nth 0 (rd :locals s)) (nth 1 (rd :locals s))) 
            (len (rd :memory s)))
        (= (nth 1 (rd :locals s)) (len (rd :memory s))))
   (= (nth 6 (rd :locals (sem-loop-8 (sem-preamble-0 s))))
      (occurlist (nth 2 (rd :locals s)) (rd :memory s)))))
\end{verbatim}

Finally, given the semantic function correctness theorems generated by 
Codewalker (namely, \texttt{SEM-PREAMBLE-0-CORRECT} and 
\texttt{SEM-LOOP-8-CORRECT}, the desired final theorem, 
depicted in Figure~\ref{FinalThm}, can be stated and proved.

\begin{figure*}
\begin{verbatim}
(defthm ll2-running-occurrences-code-=-occurlist
  (implies
   (and (hyps s)
        (program-inv s)
        (occurrences-programp s)
        (<= (+ (nth 0 (rd :locals s)) (nth 1 (rd :locals s))) 
            (len (rd :memory s)))
        (= (nth 1 (rd :locals s)) (len (rd :memory s)))
        (equal (rd :pc s) 0))
   (= (nth 6 (rd :locals (ll2 (ll2 s (clk-preamble-0 s))
                              (clk-loop-8 (ll2 s (clk-preamble-0 s))))))
      (occurlist (nth 2 (rd :locals s)) (rd :memory s))))
  :hints (("Goal" :cases ((= (len (rd :memory s)) 0) 
                          (> (len (rd :memory s)) 0)))
          ("Subgoal 2" :in-theory (enable clk-loop-8))))
\end{verbatim}
\hrulefill
\caption{Final theorem, equating the result of executing the LLVM instructions for
  the occurrences program to its abstract ``golden'' specification.}
\label{FinalThm}
\end{figure*}

\section{Related Work}

The technique of compiling to a Virtual Machine instruction set has made a
significant comeback in the past twenty years, starting with the 
JVM, and continuing with Microsoft's CIL, Android Dalvik, and 
LLVM.  Our work on verification at the virtual machine instruction set level was
inspired by J Moore's pioneering work on JVM verification \cite{MooreJVM99},
as well as Eric Smith's more recent Axe system, which was used to 
verify a number of Java cryptographic programs at the bytecode level \cite{ESmith-diss}.

Zhao \emph{et al.} \cite{Vellvm} produced several different
formalizations of operational semantics for LLVM in Coq
\cite{CoqRefMan}, noting that their intention is to produce a verified
LLVM compiler, similar to the CompCert verified compiler due to Leroy
\cite{Leroy2009} (CompCert does not utilize the LLVM intermediate
form).  The goal of Zhao \emph{et al.} was not to produce a
verification environment for LLVM bitcode, unlike the present work,
but rather to prove the correctness of compiler passes that
manipulate LLVM.  Jules Villard at Imperial College London is developing 
llStar, a formal analysis tool for LLVM bitcode.  Villard's work so
far has focused on proving 
properties of small LLVM programs that manipulate algebraic data types, 
utilizing the coreStar symbolic execution engine, separation logic, 
and SMT technology~\cite{llStar}.  LLBMC~\cite{LLBMC} is a
bounded model checker used in bug-finding for C programs that operates 
on LLVM bitcode.  Similary, KITTeL~\cite{KITTeL} performs termination
analysis on C programs by examining LLVM bitcode.  Finally, 
KLEE~\cite{KLEE} is a symbolic execution tool that operates on LLVM 
bitcode to produce coverage test cases and find bugs in C programs.

Codewalker was directly influenced by Magnus Myreen's
``decompilation into logic'' work \cite{decomp}.  It would be 
interesting to attempt to replicate the work done here using 
a combination of Myreen's decompiler and Anthony Fox's L3 
instruction set description language~\cite{L3}.  

\section{Conclusion and Future Work}

We have used Codewalker, an instruction-set-neutral
decompilation-into-logic system included with the ACL2 
theorem prover, to formally analyze C programs that have been compiled 
to the LLVM intermediate form.  Work began by defining a 
stobj-based interpreter for a subset of the LLVM instruction set, 
guided by an existing interpreter for the M1 subset of the Java 
Virtual Machine.  Several C programs, including programs to compute 
factorial, sum of array elements, and number of occurrences of a 
value in an array, were compiled to LLVM form, and hand-translated 
to an ACL2-friendly form that could be fed to the interpreter.  
Validation testing was then conducted on these programs using 
concrete inputs, before the programs were given to Codewalker for  
formal analysis.  Program-wide invariants, as well as loop invariants 
and clock measure functions, were 
defined in order to help Codewalker create semantic functions for 
program code segments.  The composition of these semantic functions 
was then proved equivalent to more abstract functions: first to a tail-recursive
form; then to a non-tail-recursive form (the equality of the 
latter two having previously been established by the
\texttt{defiteration} facility); and finally to a top-level 
non-tail-recursive ``golden'' specification.  Thus, we were 
able to prove that several sample LLVM programs implement the top-level 
specifications for those programs.

Future work will focus on using Codewalker to analyze more 
complex C functions, in particular functions that feature nested 
loops, as well as functions that employ runtime-allocated memory.  
We have successfully processed the ``straight-line'' segments for 
an LLVM insertion sort program (which features a nested loop) 
using Codewalker, but have not yet successfully 
composed the generated semantic functions into a whole 
program for further analysis.
Additionally, now that basic programs operating on unbounded integers 
have been successfully analyzed using Codewalker, a new version of 
the LLVM interpreter should be developed that can support different 
finite data word sizes, as well as the LLVM \texttt{call} and 
\texttt{ret} instructions.  Finally, we would like to apply Codewalker
to additional instruction set architectures, focusing on physical
ISAs, as opposed to virtual ISAs like LLVM.

\section{Acknowledgments}

Many thanks to J Moore for developing Codewalker.  I also wish 
to express my appreciation to  J and Matt Kaufmann for several 
emails that clarified my understanding of Codewalker's capabilities.  
Thanks also to the anonymous reviewers for their helpful comments.
Finally, I wish to acknowledge the wonderful support of my wife, Lori Hardin.

\bibliographystyle{eptcs}
\bibliography{fm}
\end{document}